\documentclass{article}
\usepackage[preprint]{spconf}
\usepackage{amsmath,amssymb,graphicx}
\usepackage[T1]{fontenc}
\usepackage{cite,booktabs,url}

\copyrightnotice{\copyright2019 IEEE}
\toappear{To appear at CAMSAP 2019}

\title{Cooperative Audio Source Separation and Enhancement Using Distributed Microphone Arrays and Wearable
Devices}
\name{Ryan M. Corey, Matthew D. Skarha, and Andrew C. Singer}
\address{University of Illinois at Urbana-Champaign}

\begin{document}
\ninept
\maketitle
\begin{abstract}
Augmented listening devices such as hearing aids often perform poorly
in noisy and reverberant environments with many competing sound sources.
Large distributed microphone arrays can improve performance, but data from remote
microphones often cannot be used for delay-constrained real-time processing.
We present a cooperative audio source separation and enhancement system
that leverages wearable listening devices and other microphone arrays
spread around a room. The full distributed array is used to separate sound sources and estimate their statistics. 
Each listening device uses these statistics to
design real-time binaural audio enhancement filters using its
own local microphones. The system is demonstrated experimentally using
10 speech sources and 160 microphones in a large,
reverberant room.
\end{abstract}


\section{Introduction}

An important application of audio signal processing is to 
help people hear better in crowded, noisy environments. Augmented
listening (AL) systems, such as hearing aids and augmented reality
headsets, alter human perception by processing sound
before it reaches the auditory system. Microphone arrays, which are used to filter signals
spatially \cite{vincent2018audio}, can improve the performance of
AL systems by separating sounds coming from
different directions  \cite{doclo2015magazine}. Large arrays can help AL systems to reduce noise more effectively,
operate with lower delay \cite{corey2018delay}, and
preserve a listener's spatial awareness \cite{marquardt2016development,koutrouvelis2018multi}.

\begin{figure}
\begin{centering}
\includegraphics{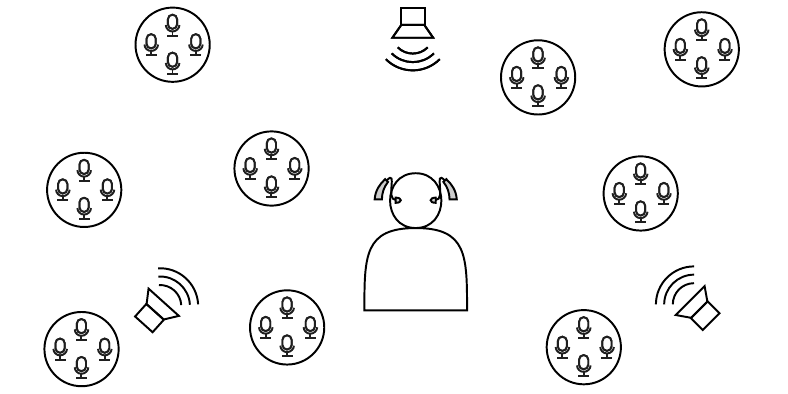}
\par\end{centering}
\caption{\label{fig:An-acoustic-environment}A room might
contain many audio devices, each with multiple microphones, that can cooperate to improve performance.}
\end{figure}

Large wearable devices with microphones spread across the body can perform better than small earpieces \cite{corey2019brtf}. Distributed arrays with sensors placed around a room could perform better still \cite{Bertrand2011}.
Microphone arrays are common in mobile and wearable devices, teleconferencing equipment, and smart-home appliances.
If these devices could be aggregated into room-scale
arrays, as shown in Fig. \ref{fig:An-acoustic-environment}, they
could dramatically improve the spatial diversity of listening systems.
There has been significant recent research interest in distributed microphone
arrays, including bandwidth-efficient distributed beamforming
algorithms \cite{bertrand2010distributed,markovich2012distributed},
distributed blind source separation methods \cite{nesta2010cooperative,himawan2010clustered,hioka2011distributed,Souden2014},
and blind synchronization to compensate for sample rate mismatch between
devices \cite{Markovich-Golan2012,Miyabe2015,Wang2016,Cherkassky2017}.

Unfortunately, distributed arrays often cannot be used directly for AL applications.
In addition to bandwidth and computational limitations, listening devices are subject to severe delay constraints: they must process sound within a few milliseconds to avoid disturbing distortion or echoes \cite{stone1999tolerable,agnew2000just}.
Even if remote microphones cannot be used directly for spatial filtering, however, they can still provide valuable information to the listening device.
Spatial filters rely on
estimates of the spatial characteristics of sound sources, such as their cross-correlation sequences or acoustic transfer functions \cite{gannot2001rtf}. Blind source
separation \cite{makino2018audio} and channel estimation \cite{Huang2006} 
methods are unreliable in large, reverberant spaces with
many competing sound sources---the very environments in which
humans most need help hearing. To reliably separate signals
and estimate parameters in challenging environments,
a single device is not enough.

\begin{figure}
\begin{centering}
\includegraphics{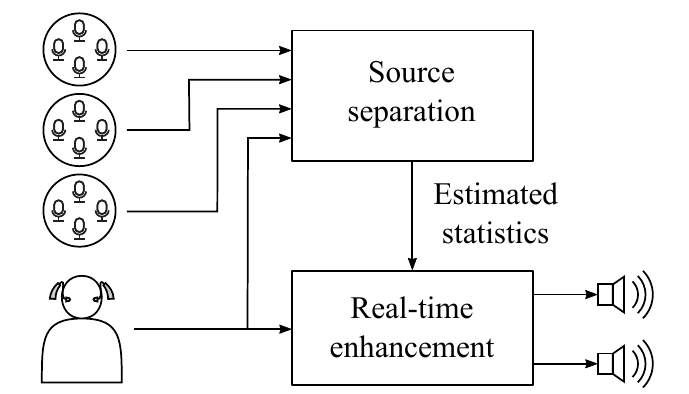}
\par\end{centering}
\caption{\label{fig:diagram}Data from many
devices is used to perform source separation and estimate signal statistics.
Listening devices perform real-time filtering using only
their own microphones.}
\end{figure}

In this work, we show how AL devices can cooperate with each other
and with other devices to improve performance in
a real-time listening enhancement task. Due to delay, computation, and bandwidth constraints, a listening device might not be able to use data
from remote devices to perform spatial filtering. Instead,
the distributed array is used to separate signals and estimate
their space-time statistics, as shown in Fig. \ref{fig:diagram}.
Each AL device uses these estimated parameters to design a real-time
multimicrophone audio enhancement filter for its local microphones.

Because the proposed system would be quite complex to implement, we
make several simplifying assumptions in this work. First, we assume
that the sources and microphones do not move. We also assume that all devices are perfectly
synchronized; for asynchronous array processing methods, see \cite{Chiba2014,corey2018async}.
To emphasize the benefits of spatial diversity from large arrays,
we ignore the temporal structure of the source signals; for distributed 
methods that leverage speech signal sparsity, see \cite{Souden2012,corey2018async,taseska2016spotforming}.
Finally, although the reverberant acoustic channel is unknown, we
assume that the number and rough locations of the sources are known.

\section{Real-Time Listening Enhancement}

Consider a room, like that in Fig. \ref{fig:An-acoustic-environment},
with $N$ sound sources and a total of $M$ microphones spread across
several devices, including at least one binaural listening device.
All $M$ microphones can be used to separate the sound sources and
estimate the parameters of the acoustic channel. However, due to delay
constraints, only $M_{\mathrm{local}}\ge2$ microphones are available
to the listening device for real-time audio enhancement. By convention,
microphone 1 is in or near the left ear and microphone 2 is in or
near the right ear.

\begin{figure}
\begin{centering}
\includegraphics{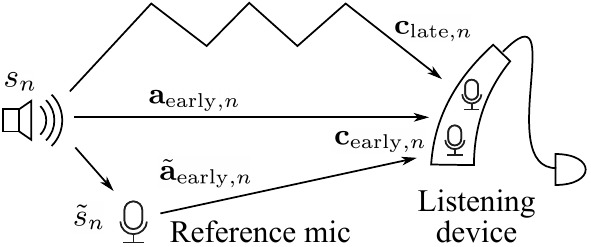}
\par\end{centering}
\caption{\label{fig:signal_model}The signal due to source $n$ is modeled as an early component and late reverberation.}

\end{figure}

Let $\mathbf{x}_{\mathrm{local}}[t]\in\mathbb{R}^{M_{\mathrm{local}}}$
be the signal received by these local microphones. It is modeled as a mixture of $N$ source images,
$\mathbf{c}_{1}[t],\dots,\mathbf{c}_{N}[t]$ and noise $\mathbf{z}[t]$,
so that 
\begin{equation}
\mathbf{x}_{\mathrm{local}}[t]={\textstyle\sum}_{n=1}^{N}\mathbf{c}_{n}[t]+\mathbf{z}[t].
\end{equation}
In a reverberant environment, each source image $\mathbf{c}_{n}[t]$
can be split into an early component $\mathbf{c}_{\mathrm{early},n}[t]$,
which includes the direct path and early reflections,
and late reverberation $\mathbf{c}_{\mathrm{late},n}[t]$, as shown in Fig. \ref{fig:signal_model} \cite{schwartz2014multi}.
There is no precise boundary between the early and late components, but it is assumed that each
early component can be modeled by an
acoustic impulse response $\mathbf{a}_{\mathrm{early},n}[k]$ 
so that
\begin{equation}
\mathbf{c}_{\mathrm{early},n}[t]={\textstyle\sum}_{k=0}^{\infty}\mathbf{a}_{\mathrm{early},n}[k]s_{n}[t-k],\label{eq:early_ir}
\end{equation}
where $s_{n}[t]$ is the signal emitted by source $n$, for $n=1,\dots,N$.

\subsection{Source remixing for augmented listening}

Human augmented listening differs from other audio enhancement applications
in two important ways. First, binaural devices must preserve spatial
perception by maintaining the interaural cues between
the left and right outputs for each source \cite{marquardt2016development,koutrouvelis2018multi}.
Second, processing delay must be no more than a few milliseconds
to avoid perceptible distortion or echoes \cite{stone1999tolerable,agnew2000just}.
This delay constraint limits the achievable performance of the system
\cite{corey2018delay}.

The AL device enhances the user's perceived auditory
scene by adjusting the levels of different sound sources,
that is, by remixing them. For simplicity, suppose that the desired
response for each source $n$ is a scalar
gain $g_{n}\ge0$. The desired output signals $y_{\mathrm{L}}[t]$
at the left ear and $y_{\mathrm{R}}[t]$ at the right ear are
\begin{align}
y_{\mathrm{L}}[t] & ={\textstyle\sum}_{n=1}^{N}g_{n}\mathbf{e}_{1}^{T}\mathbf{c}_{\mathrm{early},n}[t]\quad\text{and}\label{eq:desired}\\
y_{\mathrm{R}}[t] & ={\textstyle\sum}_{n=1}^{N}g_{n}\mathbf{e}_{2}^{T}\mathbf{c}_{\mathrm{early},n}[t],
\end{align}
where $\mathbf{e}_{m}^{T}$ is the unit vector with a 1 in position $m$.
Applying the same processing to the signals at the left
and right ears ensures that interaural cues are preserved.

\subsection{\label{subsec:causal_filtering}Delay-constrained listening enhancement}

For brevity, we henceforth restrict our attention
to the left output. A causal order-$K$ finite impulse response filter
$\mathbf{w}_{\mathrm{L}}[k]\in\mathbb{R}^{M_{\mathrm{local}}}$ produces
an output signal
\begin{equation}
\hat{y}_{\mathrm{L}}[t]={\textstyle\sum}_{k=0}^{K}\mathbf{w}_{\mathrm{L}}^{T}[k]\mathbf{x}_{\mathrm{local}}[t-k].
\end{equation}
Let the allowable delay be $\alpha$ samples so that $\hat{y}_{\mathrm{L}}[t]$
is an estimate of $y_{\mathrm{L}}[t-\alpha]$.
To derive a minimum-mean-square-error (MMSE) estimator for the desired
output, model $\mathbf{x}_{\mathrm{local}}[t]$ and $y_{\mathrm{L}}[t]$
as zero-mean wide-sense-stationary random processes. 
Let $\mathbf{r}_{xx}[k]=\mathbb{E}\left[\mathbf{x}_{\mathrm{local}}[t]\mathbf{x}_{\mathrm{local}}^{T}[t-k]\right]$ and 
$\mathbf{r}_{xy_{\mathrm{L}}}[k]=\mathbb{E}\left[\mathbf{x}_{\mathrm{local}}[t]y_{\mathrm{L}}[t-k]\right]$
be their auto- and cross-correlation functions, where $\mathbb{E}$ denotes expectation. 
Then the linear
MMSE filter that estimates $y_{\mathrm{L}}[t-\alpha]$ given $\mathbf{x}_{\mathrm{local}}[t]$
is the time-domain multichannel Wiener filter \cite{Benesty2008}
\begin{equation}
\left[\begin{matrix}\mathbf{w}_{\mathrm{L}}[0]\\
\mathbf{w}_{\mathrm{L}}[1]\\
\vdots\\
\mathbf{w}_{\mathrm{L}}[K]
\end{matrix}\right]\!=\!\left[\begin{matrix}\mathbf{r}_{xx}[0]\! & \!\mathbf{r}_{xx}[1]\! & \!\!\cdots\!\! & \!\mathbf{r}_{xx}[K]\\
\mathbf{r}_{xx}[-1]\! & \!\mathbf{r}_{xx}[0]\!\\
\vdots &  & \!\!\ddots\!\!\\
\mathbf{r}_{xx}[-K]\! &  &  & \!\mathbf{r}_{xx}[0]
\end{matrix}\right]^{-1}\!\left[\begin{matrix}\mathbf{r}_{xy_{\mathrm{L}}}[\alpha]\\
\mathbf{r}_{xy_{\mathrm{L}}}[\alpha-1]\\
\vdots\\
\mathbf{r}_{xy_{\mathrm{L}}}[\alpha-K]
\end{matrix}\right]\!.\label{eq:mwf}
\end{equation}

The cross-correlation matrix can be decomposed in terms of the source
images. From (\ref{eq:desired}), we have
\begin{align}
\mathbf{r}_{xy_{\mathrm{L}}}[k] & ={\textstyle\sum}_{n=1}^{N}g_{n}\mathbf{r}_{xc_{n}}[k]\mathbf{e}_{1},
\end{align}
where $\mathbf{r}_{xc_{n}}[k]=\mathbb{E}\left[\mathbf{x}_{\mathrm{local}}[t]\mathbf{c}_{\mathrm{early},n}^{T}[t-k]\right]$
for $n=1,\dots,N$. The listening device can easily estimate $\mathbf{r}_{xx}[k]$, but it would be difficult for it 
to estimate the source statistics $\mathbf{r}_{xc_{n}}[k]$ on its
own.

\section{Cooperative Parameter Estimation}

Although remote devices cannot be used for real-time processing,
they can be used to estimate the spatial
statistics of the sound sources. If the remote devices are spread
around and among the sound sources, then the distributed array will
have far greater spatial resolution than the listening device alone.
Furthermore, parameter estimation does not have a strict delay constraint, so the system can use several
seconds or more of audio data.

\subsection{Source separation using reference microphones}

To learn the acoustic channel parameters, we would like to estimate the source signals $s_{n}[t]$
for $n=1,\dots,N$. Unfortunately, due to permutation and scale ambiguities, we cannot directly
recover the sound produced by the source. Instead, we will estimate
each source as observed by a nearby reference microphone.

A key advantage of distributed arrays is that some devices are much
closer to some sources than others. Source separation algorithms can
exploit this spatial diversity by, for example, assigning different
sources to different devices \cite{himawan2010clustered,hioka2011distributed,Souden2014}.
Suppose that for each source $n$, there is a unique reference microphone
$m_{n}^{*}$ that is closest to it. Each reference microphone enjoys
a higher signal-to-noise ratio and direct-to-reverberant ratio than
more distant microphones for its corresponding source. Let $\tilde{s}_{1}[t],\dots,\tilde{s}_{N}[t]$
be the set of source signals as observed by their respective reference
microphones.

Let $\hat{s}_{1}[t],\dots,\hat{s}_{N}[t]$ be estimates of these reference
signals produced by a source separation algorithm. The cooperative system
of Fig. \ref{fig:An-acoustic-environment} can be used with any multimicrophone
source separation method. It remains an important open problem to
develop scalable source separation algorithms that can take full advantage
of massive-scale arrays in strongly reverberant environments with
many sources. To assess the impact of source separation on the performance
of the augmented listening system, the experiments in Sec. \ref{sec:experiments} compare
three methods with different levels of separation performance:
\begin{enumerate}
\item A baseline unprocessed estimate, which is the input mixture at the
nearest microphone to each source.
\item A blind source separation method known as independent vector analysis
(IVA), which attempts to maximize the statistical independence between
sources. We use the algorithm of \cite{Ono2011}
initialized with the nearest-microphone estimate.
\item An ideal linear MMSE filter that estimates each $\tilde{s}_{n}[t]$ 
using ground-truth acoustic channel parameters.
\end{enumerate}

\subsection{Estimation of second-order statistics}

To compute the source-remixing filter derived in Sec. \ref{subsec:causal_filtering},
we must find the second-order statistics of the early source images
$\mathbf{c}_{\mathrm{early},1}[t],\dots,\mathbf{c}_{\mathrm{early},N}[t]$.
Because the true source signals $s_{n}[t]$ are not available, we
cannot use the convolutional model (\ref{eq:early_ir}) directly.
Instead, we will use the \emph{relative }early impulse responses (REIRs) \cite{schwartz2014multi} 
$\tilde{\mathbf{a}}_{\mathrm{early},n}[k]$
with respect to the reference microphones:
\begin{equation}
\mathbf{c}_{\mathrm{early},n}[t]={\textstyle\sum}_{k=-\infty}^{\infty}\tilde{\mathbf{a}}_{\mathrm{early},n}[k]\tilde{s}_{n}[t-k],
\end{equation}
for $n=1,\dots,N$. Notice that REIRs are noncausal in general, but
if the reference microphone is close to its source, then the REIR
should closely resemble the true early impulse response. 

Because many source separation algorithms, including IVA, operate
in the time-frequency domain, it will be convenient to compute signal
statistics using the periodogram method. Let $\mathbf{X}_{\mathrm{local}}[\tau,f]$
be the short-time Fourier transform (STFT) of $\mathbf{x}_{\mathrm{local}}[t]$
and let $\hat{S}_{n}[\tau,f]$ be the STFT of $\hat{s}_{n}[t]$ for
$n=1,\dots,N$. The sample statistics are
\begin{align}
\hat{R}_{s_{n}s_{n}}[f] & =\mathrm{mean}_{\tau}\left|\hat{S}_{n}[\tau,f]\right|^{2},\quad n=1,\dots,N,\\
\hat{\mathbf{R}}_{xs_{n}}[f] & =\mathrm{mean}_{\tau}\mathbf{X}_{\mathrm{local}}[\tau,f]\hat{S}_{n}^{*}[\tau,f],\quad n=1,\dots,N,\\
\hat{\mathbf{R}}_{xx}[f] & =\mathrm{mean}_{\tau}\mathbf{X}_{\mathrm{local}}[\tau,f]\mathbf{X}_{\mathrm{local}}^{H}[\tau,f].
\end{align}
Note that these sample statistics are only correct if the signals
are wide-sense stationary, which is not true in practice. By using
the long-term average statistics, we ignore the temporal nonstationarity
of the source signals and rely on spatial diversity alone.

If the sources and noise are uncorrelated with each other, then we
can estimate the discrete-frequency relative transfer function of
$\mathbf{c}_{n}[t]$ with respect to microphone $m_{n}^{*}$ as
\begin{equation}
\hat{\mathbf{A}}_{n}[f]=\hat{\mathbf{R}}_{xs_{n}}[f]\hat{R}_{s_{n}s_{n}}^{-1}[f],\quad n=1,\dots,N.
\end{equation}
The relative early transfer function $\hat{\mathbf{A}}_{\mathrm{early},n}[f]$
is obtained by time-domain windowing. The length of this window is
a tunable parameter that, based on our experiments, does not appear
to have a strong impact on objective performance.

The estimated cross-spectra between the mixture and images are
\begin{equation}
\hat{\mathbf{R}}_{xc_{n}}[f]=\hat{\mathbf{R}}_{xs_{n}}[f]\hat{\mathbf{A}}_{\mathrm{early},n}^{H}[f],\quad n=1,\dots,N.
\end{equation}
The correlation functions required to compute the remixing filter
are obtained by taking the inverse discrete Fourier transform of $\hat{\mathbf{R}}_{xx}[f]$
and $\hat{\mathbf{R}}_{xc_{n}}[f]$ for $n=1,\dots,N$. 

\section{Experiments}
\label{sec:experiments}

\subsection{Experimental setup}

\begin{figure}
\begin{centering}
\includegraphics[width=2cm]{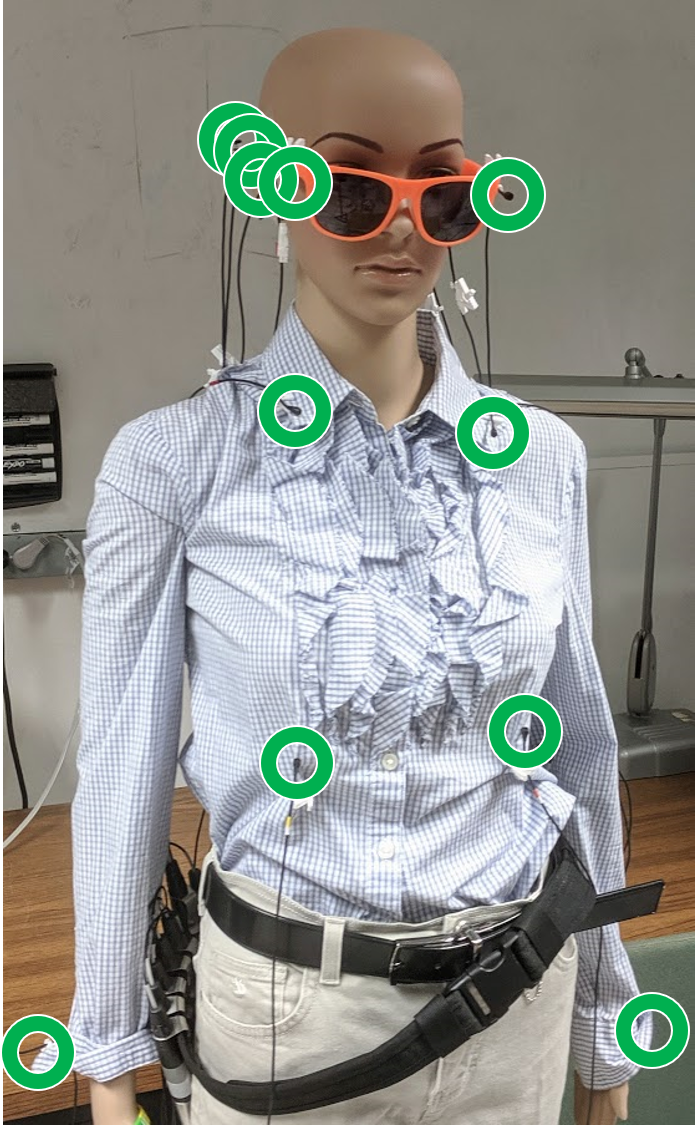} \hfill \includegraphics{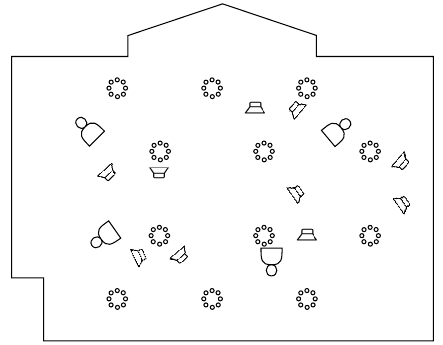} \hfill 
\includegraphics[width=1.75cm]{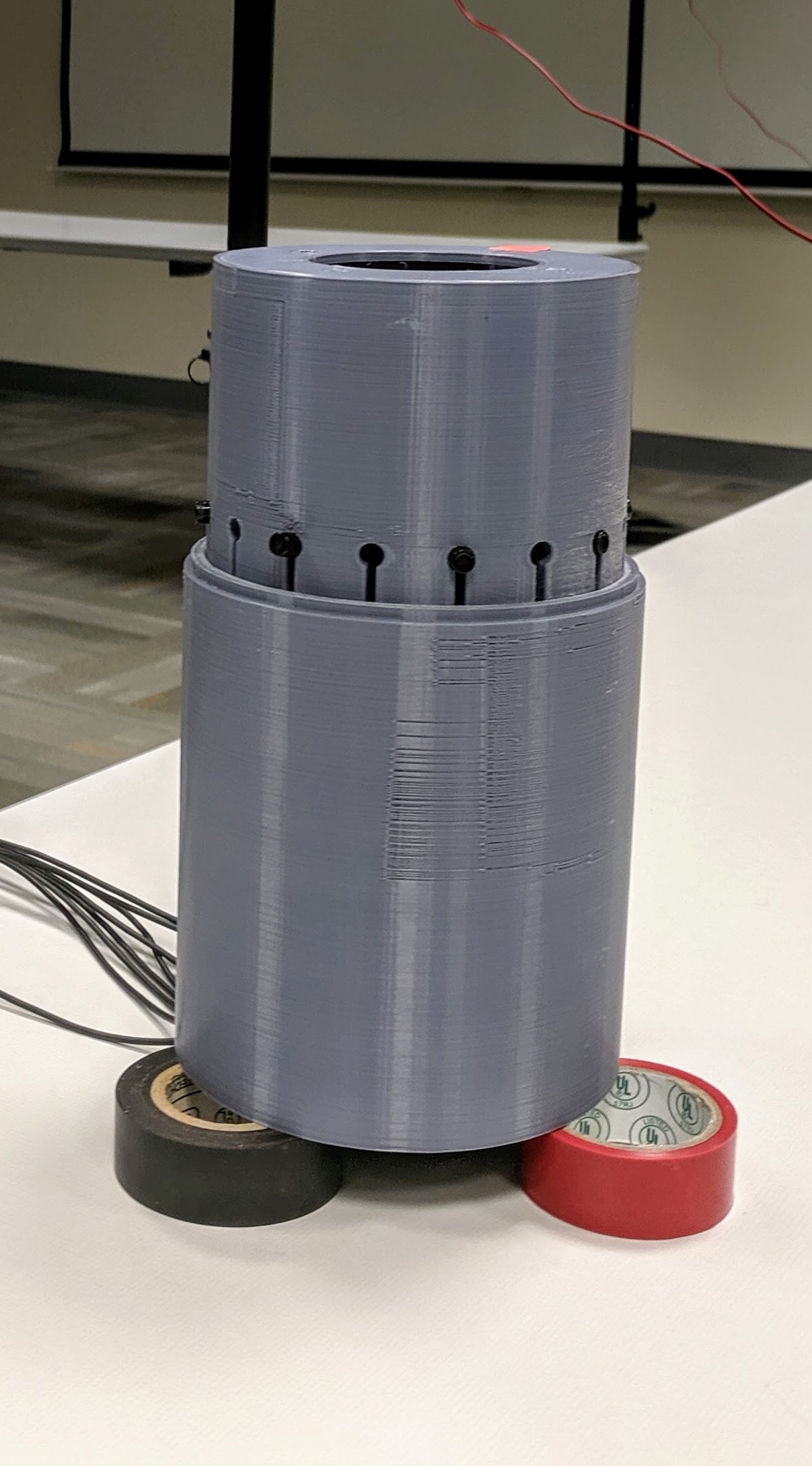}
\par\end{centering}
\caption{\label{fig:layout}Four wearable arrays (left), twelve tabletop arrays (right), and ten loudspeakers were spread around
a large room.}
\end{figure}

To demonstrate the proposed cooperative source
separation and enhancement system in a challenging real-world environment,
an experiment was conducted using 10 loudspeakers and a total of
160 omnidirectional microphones in a large (9 m by 13 m), strongly reverberant conference room ($T_{60}\approx780$
ms), shown in Fig. \ref{fig:layout}. Twelve enclosures, designed to resemble voice-assistant speakers, held eight
microphones each in a circular pattern with diameter 10 cm. The remaining
64 microphones were in wearable arrays on four plastic mannequins.
Each had one microphone near each ear canal,
four in behind-the-ear earpieces, two
on eyeglasses, and eight on a shirt. Due to equipment limitations, recordings were captured using sixteen
microphones at a time and the devices were moved between recordings
while the ten loudspeakers remained fixed.

The loudspeakers played quasi-anechoic speech samples from ten
talkers in the VCTK corpus \cite{Veaux2017}. To quantify the source
separation and enhancement performance of the system, each sound source
was played back and recorded separately to capture the source 
images, which can be added together to form test mixtures. 
The dataset is available on the Illinois Data Bank \cite{corey2019massive}.

The closest microphone to each source was selected as the reference
for source separation using the baseline, blind, and ideal methods described above. Source separation and parameter estimation were performed using
16 seconds of audio data. Different
16-second speech clips from the same talkers were used to evaluate the resulting listening enhancement filters. 
The filters have a target delay of 16 ms and an impulse response length
of 128 ms. The length of the REIRs used to model the target sources' acoustics was 32 ms.

To consistently quantify performance, the enhancement filters (\ref{eq:desired})
were designed to isolate a single source at a time, so that $g_n = 1$ for target source $n$ and 0 for all others. 
A total of $8N$ single-target enhancement filters were designed, one for each source and each ear.
The output signal-to-noise ratio (SNR) for a source separation or
enhancement filter $\mathbf{w}$ designed to isolate source $n$
is
\begin{equation}
\text{SNR}_{n}=10\log_{10}\frac{\sum_{t} \left(\sum_{k} \mathbf{w}^T[k]\mathbf{c}_{n}[t-k]\right)^{2}}{\sum_{t}\left(\sum_{p\ne n}\sum_{k}\mathbf{w}^T[k] \mathbf{c}_{p}[t-k]\right)^{2}}.
\end{equation}

Note that separately recording and combining source images has
the effect of amplifying ambient noise. For qualitative
evaluation of enhancement results under more realistic conditions, the
experiment was repeated using a simultaneous recording of all ten
sources;  binaural samples are available at the first
author's website\footnote{\url{http://ryanmcorey.com/demos}}.

\subsection{Experimental results}

The separation system was evaluated with several array
configurations: the 10 reference microphones alone, the 4 wearable arrays, 
the 12 tabletop arrays,
and all 160 microphones together. Table \ref{tab:IVA-Source-Separation}
shows the median SNR improvement of the estimated reference signal $\hat{s}_n[t]$
compared to the unprocessed nearest-microphone signal for $N=4$, $7$, and $10$ sources. While
IVA performs better than the baseline, especially for small
$N$, it does not scale well with increasing array size and there
is a large gap between its performance and that of the ideal filter. 
Notice that the wearable arrays outperform the tabletop arrays despite having fewer total microphones; the acoustically opaque body improves the spatial diversity of these arrays \cite{corey2019brtf}.

\begin{table}
\begin{centering}
\setlength{\tabcolsep}{5pt}
\begin{tabular}{cccccccccc}
\toprule 
 &  & \phantom{a}  & \multicolumn{3}{c}{Ideal } & \phantom{a}  & \multicolumn{3}{c}{IVA} \\
\cmidrule{4-6} \cmidrule{8-10}
Array & $M$ &  & $N=4$ & 7 & 10 &  & $N=4$ & 7 & 10 \\
\midrule 
Reference & 10 &  & 13 & 12 & 10 &  & 5 & 4 & 3 \\
Wearable & 64 &  & 25 & 26 & 23 &  & 8 & 7 & 3 \\
Tabletop & 96 &  & 23 & 23 & 21 &  & 7 & 6 & 5 \\
All mics & 160 &  & 23 & 24 & 23 &  & 8 & 7 & 6 \\
\bottomrule 
\end{tabular}
\end{centering}
\caption{\label{tab:IVA-Source-Separation}Median source separation performance measured by SNR improvement, in dB, between the estimated reference signal and the noisy signal at the reference microphone.}

\end{table}

\begin{figure}
\begin{centering}
\includegraphics{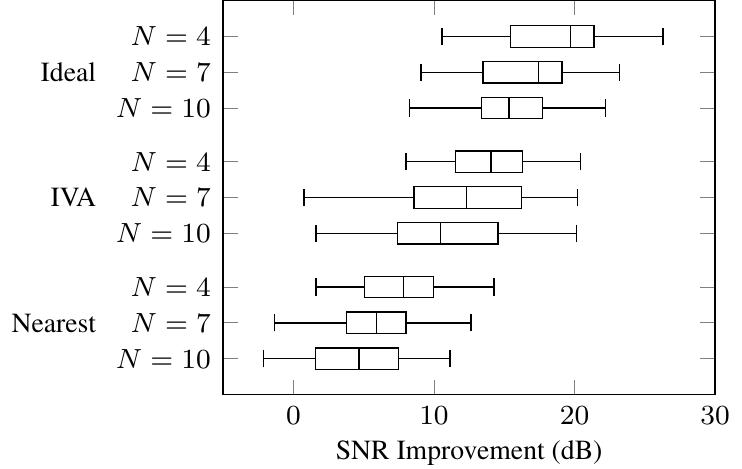}
\par\end{centering}
\caption{\label{fig:Listening-enhancement-performanc}Listening enhancement
performance for different numbers of sources and different separation
methods. All 160 microphones were used for separation. The boxes show
the quartile statistics over all $8N$ source-output pairs.}
\end{figure}

Figure \ref{fig:Listening-enhancement-performanc} shows the SNR improvement
of the AL device output compared to the unprocessed input at each ear.
Within each experiment, lower SNR improvements generally correspond
to distant source-listener pairs and larger improvements are for
nearby source-listener pairs. For example, for the 7-source mixture
using IVA parameters, the listener in the upper left corner
in Fig. \ref{fig:layout} achieved an 18 dB SNR improvement for the
directly adjacent source but only a 1 dB improvement for the source
in the opposite corner of the room. 

Listening enhancement performance appears to depend strongly on source
separation performance. There is
a roughly 5 dB performance difference between the filters designed
from the unprocessed reference microphone signals and those designed
from the IVA estimates, showing that cooperative source separation 
did improve the performance of the individual
AL devices. There is also a 5 dB difference between
the IVA-based filters and those designed from ideal estimates, showing
that there is room for improvement in distributed source separation. 

\begin{figure}
\begin{centering}
\includegraphics{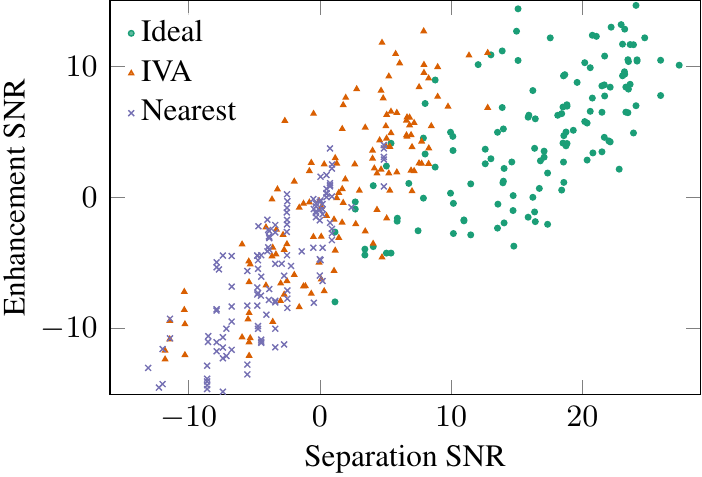}
\par\end{centering}
\caption{\label{fig:snr_comparison}Enhancement SNR as a function
of separation SNR. Each point represents one source-to-ear
filter. Results are sampled from all array and source configurations
in Table \ref{tab:IVA-Source-Separation}.}
\end{figure}

To further illustrate the relationship between source separation and
audio enhancement performance, Fig. \ref{fig:snr_comparison} shows
enhancement SNR as a function of separation SNR for individual
source-microphone pairs. For the two non-ideal separation methods,
there appears to be a roughly linear relationship: every 1 dB improvement
in the separation SNR provides about 1 dB improvement in enhancement
SNR. The ideal unmixing filter shows diminishing returns above around
10 dB. The vertical spread in the figure appears to be due to different
distances between sources and listeners: nearby listeners achieve
larger enhancement gains compared to distant listeners for the same
source estimate.

\section{Conclusions}

The experiments presented above show that a distributed room-scale
array can help listening devices to perform useful audio enhancement
in challenging reverberant environments where source separation would
otherwise be difficult or impossible. In contrast to other distributed methods,
data from the distributed array is not used for real-time filtering
but for parameter estimation. While a 16-microphone wearable array
can provide strong enhancement performance on its own, it cannot reliably
estimate the space-time statistics of the sources. The cooperative processing system can.

Because the
enhancement filter is designed to match the estimated reference signal,
its performance depends strongly on that of the source separation algorithm. Further research
is required to find separation methods that can take advantage of massive
spatial diversity to reliably separate large numbers of sources in
strongly reverberant environments. With these improvements, the proposed
cooperative audio separation and enhancement system will allow augmented listening
devices to leverage all the microphones in a room, helping users to
hear clearly in even the most challenging situations.

\bibliographystyle{ieeetran}
\bibliography{../master_references}

\end{document}